\documentclass[12pt]{article} 
\title{Systemic risk through contagion in a core-periphery structured banking network}

\author{Oliver Kley\thanks{Centre for Mathematical Sciences, Technische Universit\"at M\"unchen,  85748 Garching, Boltzmannstrasse 3, Germany, email: \{oliver.kley\,,\,cklu\}@tum.de}
\and
Claudia Kl\"uppelberg\footnotemark[1]
\and
Lukas Reichel\thanks{Institute of Insurance Economics, University of St. Gallen,  9000 St. Gallen, Tannenstrasse 19, Switzerland, email: lukas.reichel@unisg.ch}
}

\usepackage{amssymb}
\usepackage{amsfonts}
\usepackage{amsmath}
\usepackage{amsthm}
\usepackage{graphicx}
\usepackage{caption}
\usepackage{subcaption}
\usepackage[usenames, dvipsnames]{xcolor}
\usepackage{verbatim}
\usepackage{dsfont}
\usepackage{color}
\usepackage[skip=2pt,font=footnotesize]{caption}
\numberwithin{equation}{section}

\newtheorem{theorem}{Theorem}[section]
\newtheorem{lemma}[theorem]{Lemma}
\newtheorem{remark}[theorem]{Remark}
\newtheorem{example}[theorem]{Example}
\newtheorem{proposition}[theorem]{Proposition}
\newtheorem{definition}[theorem]{Definition}
\newtheorem{assumption}[theorem]{Assumption}
\newtheorem{corollary}[theorem]{Corollary}
\newtheorem{fig}[theorem]{Figure}

\newcommand{\bthe}{\begin{theorem}}
\newcommand{\ethe}{\end{theorem}}

\newcommand{\ben}{\begin{enumerate}}
\newcommand{\een}{\end{enumerate}}

\newcommand{\bit}{\begin{itemize}}
\newcommand{\eit}{\end{itemize}}

\newcommand{\beq}{\begin{equation}}
\newcommand{\eeq}{\end{equation}}

\newcommand{\ble}{\begin{lemma}}
\newcommand{\ele}{\end{lemma}}

\newcommand{\bde}{\begin{definition}\rm}
\newcommand{\ede}{\halmos\end{definition}}

\newcommand{\bco}{\begin{corollary}}
\newcommand{\eco}{\end{corollary}}

\newcommand{\bpr}{\begin{proposition}}
\newcommand{\epr}{\end{proposition}}

\newcommand{\brem}{\begin{remark}\rm}
\newcommand{\erem}{\halmos\end{remark}}

\newcommand{\bproof}{\begin{proof}}
\newcommand{\eproof}{\end{proof}}

\newcommand{\bexam}{\begin{example}\rm}
\newcommand{\eexam}{\halmos\end{example}}

\newcommand{\beass}{\begin{assumption}}
\newcommand{\eass}{\end{assumption}}

\newcommand{\bfi}{\begin{fig}}
\newcommand{\efi}{\end{fig}}

\newcommand{\btab}{\begin{tab}}
\newcommand{\etab}{\end{tab}}

\newcommand{\beao}{\begin{eqnarray*}}
\newcommand{\eeao}{\end{eqnarray*}\noindent}

\newcommand{\beam}{\begin{eqnarray}}
\newcommand{\eeam}{\end{eqnarray}\noindent}

\newcommand{\barr}{\begin{array}}
\newcommand{\earr}{\end{array}}

\newcommand{\bdis}{\begin{displaymath}}
\newcommand{\edis}{\end{displaymath}\noindent}

\def\E{{\mathbb E}}
\def\R{{\mathbb R}}

\def\calf{{\mathcal{F}}}

\newcommand{\tto}{t\to\infty}

\newcommand{\al}{{\alpha}}

\newcommand{\si}{{\sigma}}

\newcommand{\vp}{\varphi}
\newcommand{\eps}{\varepsilon}

\newcommand{\var}{{\rm Var}}

\newcommand{\ov}{\overline}

\newcommand{\halmos}{\quad\hfill\mbox{$\Box$}}  

\allowdisplaybreaks[4]

\begin{document}


\maketitle

\begin{abstract}
We contribute to the understanding of how systemic risk arises in a network of credit-interlinked agents.
Motivated by empirical studies we formulate a network model which, despite its simplicity, depicts the nature of interbank markets better than a homogeneous model.
The components of a vector Ornstein-Uhlenbeck process living on the vertices of the network describe the financial robustnesses of the agents. For this system, we prove a LLN for growing network size leading to a propagation of chaos result. 
We state properties, which arise from such a structure, and examine the effect of inhomogeneity on several risk management issues and the possibility of contagion. 
\end{abstract}

\noindent
\begin{tabbing}
{\em AMS 2010 Subject Classifications:} \=  60K35\,; 
60H30\,; 
91B30\,. 
\\
{\em JEL Classification:} \= G18\,; G21\,; G23\,.
\end{tabbing}


\noindent
{\em Keywords:  core-periphery bank model, financial contagion, inhomogeneous graph, interacting particles, systemic risk}


\section{Introduction}\label{s1}

Interbank lending patterns and financial contagion have been in the focus of central banks and regulators already before, but predominantly since the financial crisis has started in 2007, succeeded by the government debt crisis in Europe.
Consequently, there is a number of empirical studies performed by central banks dealing with issues of contagion; see e.g. \cite{ DN, Furfine, Mistrulli, Mueller, SM, UW, Wells}. However, as indicated in Mistrulli~\cite{Mistrulli}, data limitations - especially not or only partly available bilateral exposures between agents - often enforced the use of the so-called maximum-entropy method. 
This method actually rules out structural information about the market, while assuming that each bank lends to all others, possibly leading to over- or underestimation of contagion (cf. Mistrulli \cite{Mistrulli} for a discussion and data analysis). 
With respect to structural properties of interbank markets, the empirical studies \cite{bossetal} for the Austrian market,  \cite{soramaki:2007}  for the Fedwire interbank payment network, \cite{CMS} for the Brazilian market, and \cite{craig:2014} for the German market share the same finding: 
There is a small number of highly connected big banks acting as financial intermediaries for a large number of smaller banks, which mostly do not interact directly. 
Our approach weakens the homogeneity assumptions by suggesting a more realistic model allowing for top-tier and lower-tier banks. 

Several studies model the financial market as a random graph.  
Financial contagion in  the market is then treated by investigating and simulating a discrete bankruptcy cascade, which is initiated by some triggering mechanism like a first passage event;
see e.g.  \cite{cont:resilience, battiston:2012, gaikapadia,hurdgleeson}.  
Another approach is based on mean field models of interacting systems of diffusions as used in physics to model the evolution of particles.
This yields a homogeneous financial market; see e.g.~\cite{FS, GPY, garnieretal}.
Our approach extends such models two-fold. Firstly, we replace the driving Brownian motion (BM) by a L\'evy process, which does not require new techniques. 
Secondly and more important, we modify the model away from homogeneity to the  above mentioned two-tier market structure, 
and derive for this a new limit theorem.  
We thus enter a new line of research that allows for more flexibility in modeling the robustness of the financial market and its agents.

Our paper is organised as follows. 
We introduce the two-tier financial market model in Section~\ref{s2} and present subsequently the robustness process for all agents in the market. 
In Section~\ref{s3} we prove a LLN for the new financial system, which may be interpreted as a propagation of chaos result.
The limit model is further studied and used for the sake of systemic risk assessment in Section~\ref{s4}.
We investigate the systemic risk of the market in terms of the standard deviation risk and the inverse first passage time risk.
We examine in particular the effect of individual risk management decisions on the risk of contagion. The paper concludes with an outlook on future research in Section~\ref{s5}.

\section{The contagion model}\label{s2}

\subsection{Market model}\label{s2_1}

We  model a credit interbank market as a {\em weighted directed graph} $G^w = (V, E, W)$  with a finite number of vertices $|V|=N$. This is a common approach, see e.g. \cite{cont:resilience, battiston:2012}. 
While the set of vertices $V$ represents the agents in the market, information about their bilateral credit relationships is  encoded in the set $E$ of edges. In the interbank market each agent $i \in V$ manages a credit portfolio $E_i\subset E$, where $(i,j)\in E_i$ holds if and only if agent $j$ is a debtor of agent $i$. 
We agree upon that no agent lends to herself, i.e. $(i,i)\notin E_{i}$ for all $i \in V$ and denote by $V_{out}(i)=\{j\in V : (i,j)\in E\}$ the set of debtors of agent $i$. Correspondingly, $d_{out}(i)=|V_{out}(i)|$ indicates the out-degree of agent $i$ being the number of issued credits by agent $i$ in our context. The set $W$ endows every edge $(i,j) \in E$ with an individual weight $w_{ij}\in (0,1]$, that is, the credit issued from agent $i$ to agent $j$ corresponds to $w_{ij} \cdot 100$ \% of the total credit amount agent $i$ has issued to the overall interbank market. Consequently,  
\begin{eqnarray*}
\sum_{j\in V_{out}(i)} w_{ij} = 1.
\end{eqnarray*} 
In the homogeneous graph of \cite{battiston:2012} each agent issues credits to exactly $0\le k < N$ other agents from the same market so that $d_{out}(i) = k$ for all $i \in V$. Moreover, the credit weights are assumed to be uniformly $w_{ij} = \frac{1}{k}$ for all $j\in V_{out}(i)$ and 0 otherwise. 

We extend this model to an inhomogeneous graph in the sense that we allow for two types of agents, the {\em core banks} and the {\em periphery banks}. 
There is empirical evidence for a two-tiered structure of interbank markets: top-tier banks and lower-tier banks, cf.  \cite{bossetal,CMS,soramaki:2007, UW} and in particular \cite{craig:2014}, which develops a more specific core-periphery network model. 
In a simplified purely tiered network top-tier (core) banks can potentially lend to and borrow from any bank in the network, while lower-tier (periphery) banks exclusively interact with top-tier banks but not with banks from their own tier. 

We underlay this core-periphery interbank market with the following specifying assumptions:
\begin{itemize}
\item The interbank market is partitioned into a set of core banks $C$ and a set of periphery banks $P$; i.e., $V=C \cup P$.
\item  The set of debtor banks $V_{out}^C(i)$ of a core bank $i \in C$ can be partitioned into the two subsets
\begin{eqnarray*}
V^{CC}_{out}(i) &:=& \{j \in C: (i,j) \in E \}\quad\mbox{and}\quad
V^{CP}_{out}(i) \, := \, \{j \in P: (i,j) \in E \}.
\end{eqnarray*}
Analogously, for a periphery bank $i \in P$ we set
\begin{eqnarray*}
V^{PP}_{out}(i)&:=& \{j \in P: (i,j) \in E \}\quad\mbox{and}\quad
V^{PC}_{out}(i) \, :=\,  \{j \in C: (i,j) \in E \}.
\end{eqnarray*} 
\item The banks (nodes) have the following out-degree structure:
$$\barr{rclll}
d_{out}^{CC}(i) &=& k_{CC} \hspace{0.5cm}\mbox{ with } &  0\leq k_{CC}\leq |C|-1 \hspace{0.3cm} & \mbox{ for } i \in C,
\newline \\
d_{out}^{CP}(i) &=& k_{CP} \hspace{0.5cm}\mbox{ with } &  0\leq k_{CP}\leq |P|\hspace{0.3cm} & \mbox{ for } i \in C,
\newline \\
d_{out}^{PP}(i) &=& k_{PP} \hspace{0.5cm}\mbox{ with } &  0\leq k_{PP}\leq |P|-1\hspace{0.3cm} & \mbox{ for } i \in P,
\newline \\
d_{out}^{PC}(i) &=& k_{PC} \hspace{0.5cm}\mbox{ with } &  0 \leq k_{PC}\leq |C|\hspace{0.3cm} & \mbox{ for } i \in P.
\earr$$
\item  Each agent acts as a creditor and issues credits to other agents from the same market:
\begin{eqnarray*}
\max\{k_{CC}, k_{CP}\} &>&0\quad\mbox{and}\quad
\max\{k_{PP}, k_{PC}\} \, > \, 0.
\end{eqnarray*}
\end{itemize}
The \textit{adjacency matrix} $A = (a_{ij})_{i,j = 1}^N$ indicates the bilateral credit relationships between the agents of the network by entries of ones and zeros, more precisely,
\begin{eqnarray}\label{eq:aMatrix}
a_{ij} = \begin{cases}1 & \text{if } (i,j) \in E \\
						0 & \text{else.}
\end{cases}
\end{eqnarray}
In view of the two-tiered structure of the market, a block model can be employed, which is a common approach in social network analysis; cf. \cite{wasserman1994social}.
In our case the adjacency matrix $A$ is a block matrix composed of 4 blocks corresponding to the core and periphery decomposition of $V$:
\begin{eqnarray}\label{eq:block}
A = \begin{pmatrix} 
CC & CP \\ 
PC & PP 
\end{pmatrix}.
\end{eqnarray}
The block $CC$ having dimension $|C| \times |C|$ lists the credit relationships among the core banks, the $|P| \times |P|$ block $PP$ provides the information about the relationships among the periphery banks and the blocks $PC$ and $CP$ cover the exchange of credits between core and periphery, respectively.

The weighted adjacency matrix $A^w = (a_{ij}^w)_{i,j = 1}^N$ is defined  through
\begin{eqnarray*}
a_{ij}^w = \begin{cases}w_{ij} & \text{if } (i,j) \in E \\
						0 & \text{else.} \end{cases}
\end{eqnarray*}

\bexam\label{ex:2type}[Craig and von Peter~\cite{craig:2014}]
Here the blocks are specified as follows:
\begin{itemize}
\item  $CC$ is a matrix of ones exceptional the zero diagonal: all core banks issue credits to all other core banks;
\item $PP$ is a matrix of zeros: periphery banks issue no credits among each others;
\item  $CP$ is row regular, that is, each row has at least one 1: each core bank issues credits to at least one periphery bank;
\item  $PC$ is column regular, that is, each column is covered by at least one 1: at least one periphery bank issues a credit to one of the core banks.
\end{itemize}
\eexam

\subsection{Financial robustness}\label{ss22}

Following \cite{battiston:2012, FS}, we endow each agent $i \in V$ in our network by a measure called {\em financial robustness} which quantifies an agent's financial constitution over time. In the following we specify this measure as a continuous-time stochastic process, where all stochastic quantities will be defined on a probability space $(\Omega,\calf,P)$. In our specification we suppose that the behavior of the financial robustness is related to two sources: 
On the one hand, an agent's robustness depends on the robustness of its debtors. If the debtors' robustness is low, an agent has to face higher counterparty risk and, thus, its robustness will suffer as well. 
On the other hand, the robustness will also be affected by any non-interbank market investment. 
We model this by a vector Ornstein-Uhlenbeck process $\rho$ given as solution of the vector stochastic differential equation (SDE)
\begin{eqnarray}\label{eq:moup}
d \rho_t = (A^w - I_{N \times N})\rho_t d t +  dL_t,\quad t\ge0,
\end{eqnarray}
where $L=(L_t)_{t\ge 0}$ denotes an $N$-dimensional mean 0 L\'evy process with finite variance. 
The component $(A^w - I_{N \times N})\rho_t$ models the interdependence resulting from the agents' interbank market activity, whereas the L\'evy process covers the impact from external market sources. 
By incorporating $A^w$ the robustness process is explicitly addressing the network structure. For our purposes this network structure is kept constant over time, which is in line with the findings of \cite{craig:2014} about the structural stability of the German interbank market. 

The following result gives the solution of the SDE \eqref{eq:moup} and the second order moment structure.
\bpr\label{prop:OUprocess}
For the SDE \eqref{eq:moup} with $\rho_t=(\rho_t^1,\ldots,\rho_t^N)$ and initial vector $\rho_0=(\rho_0^1,\ldots,\rho_0^N)$  the following assertions hold.\\[2mm]
(a) \,
The SDE has a unique explicit solution given by
\begin{eqnarray}\label{rhomult}
\rho_t = \exp\left[t(A^w - I_{N \times N})\right]\rho_0 +  \int_0^t \exp\left[(t-s)(A^w - I_{N \times N})\right]  dL_s,\quad t\ge0,
\end{eqnarray}
with the matrix exponential $\exp[X] := \sum_{m = 0}^{\infty} \frac{1}{m!} X^m$, and $I_{N \times N}$ is the unit matrix.\\[2mm]
(b) \,The mean of the process is given by
$$E\left[\rho_t\mid \rho_0\right] = \exp\left[t(A^w - I_{N \times N})\right]\rho_0,\quad t\ge 0.$$
(c) \, 
For every $t,t^{\prime}>0$ the covariance matrix function is given by
\begin{eqnarray*}
Cov\left[\rho_t, \rho_{t^{\prime}}\right] = \Sigma \int\limits_0^{\min(t,t^\prime)}\exp\left[(t-s)(A^w - I_{N \times N})\right]\exp\left[(t^{\prime}-s)(A^w - I_{N \times N})^\top\right]ds,
\end{eqnarray*}
where $\Sigma$ is the diagonal variance matrix of $L_1$.
\epr
For information and details on L\'evy processes we refer to~\cite{Applebaum04} or~\cite{Sato}.

\section{Financial robustness in large networks}\label{s3}

If we pick out one row of Eq.~\eqref{eq:moup}, then the financial robustness of agent $i\in V$ follows the dynamic
\begin{eqnarray}\label{eq:rhoi}
d\rho^i_t = \Big(\sum_{j\in V_{out}(i)} w_{ij}\rho^j_t-\rho^i_t\Big)dt + dL^i_t,\quad i\in V.
\end{eqnarray}
As described above the drift term adjusts the process towards the mean robustness of agent $i$'s debtors. Note that the mean is calculated over $\rho^j_t$ with $j\neq i$, hence this ensemble mean is independent of the driving process $L^i$.

When all weights $w_{ij}$ are chosen to be equal and the driving process is a Brownian motion, then this is a classical example in physics for interacting particle systems going back to McKean.
We extend McKean's mean field example of interacting diffusions to the inhomogeneous system \eqref{rhomult} driven by independent L\'evy processes. 

We choose the weights $w_{ij}$ based on the following market assumption, which are in line with with a perfectly tiered interbank market as considered in \cite{craig:2014}.
All core banks interact with each other and every periphery bank is creditor and debtor to every core bank. 
For the periphery banks, any credit relationship among them is excluded.
Then the SDE \eqref{eq:rhoi} becomes 
\beam\label{eq:rhoicore}
d\rho^i_t &=& \Big(\sum_{j\in C\setminus{\{i\}}} w_{ij} \rho^j_t + \sum_{k\in P} w_{ik}\rho^k_t - \rho_t^i\Big) dt + \si_C dL^i_t,\quad i\in C,\\
d\rho^k_t &=& \Big(\sum_{i\in C} w_{ki} \rho^i_t  - \rho_t^k\Big) dt + \si_P dL^k_t,\quad k\in P,\label{eq:rhoiperiphery}
\eeam
where all L\'evy processes are independent with mean $E[L_1^i]=0$, and  standardized second moment $E[(L_1^i)^2]=1$ for $i\in V$.
{The constants $\si_C,\si_P>0$ model the standard deviations of the core and periphery banks respectively. The L\'evy processes $L_i$ are for all $i\in C$ identically distributed, as well as the $L_k$ for all $k\in P$.}
Moreover, we assume the following simple scenario for the weights.
For all $i\in C$ we assume that $w_{ij} = \frac{1-\eps}{|C|-1}$ for $j\in C\setminus{\{i\}}$ and for some $\eps\in (0,1)$, and also that $w_{ik} = \frac{\eps}{|P|}$ for $k\in P$, so that  $\sum_{j=1,\, j\neq i}^N w_{ij}=1$. 
For all $k\in P$ we assume that $w_{ki}=\frac1{|C|}$ for all $i\in C$ and $w_{ki}=0$ for all $i\in P$.
Then \eqref{eq:rhoicore} and \eqref{eq:rhoiperiphery} read as
\beam\label{eq:rhocoreequalweights}
d\rho^i_t &=& \Big(\frac{1-\eps}{|C|-1}\sum_{j\in C\setminus{\{i\}}} \rho^j_t + \frac{\eps}{|P|}\sum_{k\in P} \rho^k_t - \rho_t^i\Big) dt + \si_C dL^i_t,\quad i\in C,\\
d\rho^k_t &=& \Big(\frac{1}{|C|}\sum_{i\in C} \rho^i_t  - \rho_t^k\Big) dt + \si_P dL^k_t,\quad k\in P.\label{eq:rhoperipheryequalweights}
\eeam

This is a coupled system, where the robustness of the core banks is influenced by the mean robustness of all other core banks and the mean robustness of all periphery banks. 
The robustness of the periphery banks, on the other hand, is influenced by that of the core banks only.
We prove a LLN for the empirical distributions given by the weighted sums when the system becomes large; i.e. for $N\to\infty$.

\bthe\label{meanfieldlimit}
Assume the core-periphery model \eqref{eq:rhocoreequalweights} and \eqref{eq:rhoperipheryequalweights} with independent driving L\'evy processes, { which are identically distributed for all $i\in C$ and all $k\in P$, respectively.}
Define the limit system by the dynamics
\beam\label{limitC}
 d\ov\rho^i_t
&=& \Big( (1-\eps) E[\ov\rho_t^C] +\eps  E[\ov\rho_t^P] -\ov\rho^i_t\Big) dt + \si_C dL^i_t,\quad i\in C,\\
d\ov\rho^k_t &=& \Big( E[\ov\rho_t^C]-\ov\rho^k_t\Big)dt + \si_P dL^k_t,\quad k\in P,\label{limitP}
\eeam
where  $E[\ov\rho_t^C] = \int _\R y \mu_t^C(dy)$ and $E[\ov\rho_t^P] = \int _\R y \mu_t^P(dy)$; i.e. $\mu_t^C$ is the distribution of $\ov\rho^i_t$ for all $i\in C$ and $\mu_t^P$ that of $\ov\rho^k_t$ for all $k\in P$.
Take  the same driving L\'evy processes as above and the same initial conditions  $\rho^i_0 = \ov\rho^i_0$ for $i \in V$, independent of all L\'evy processes.
Denote $|x-y|^*_T := \sup_{t\le T} |x_t-y_t|$. Then for every  $T>0$, $|C|/|P|\le M<\infty$, and a  constant $c>0$ independent of $|C|$,
\beam\label{eq:meanfieldlimit}
\sqrt{|C|} \, E[|\rho^i-\ov\rho^i|^*_T] \leq  c < \infty,\quad i\in V.
\eeam
\ethe

\bproof
For the proof we adapt the arguments of the proof of Theorem~1.4 of Sznitman~\cite{Sznitman} to the inhomogenous system.
First note that for $k\in P$
\beao
\rho_t^k-\ov \rho_t^k &=& \int_0^t  ds \Big\{ \Big(\frac{1}{|C|} \sum_{i\in C} (\rho^i_s - E[\ov\rho_s^C])\Big)  - (\rho_s^k   -\ov\rho_s^k )\Big\}.
\eeao
Summing this equality over all $k\in P$, 
and using the fact that $\rho^k$ and $\ov\rho^k$ for $k\in P$ are equally distributed, respectively, we obtain for $l\in P$ and some $K>0$ ($K$ always denotes some positive constant, whose value may vary from line to line) { by taking the modulus under the Lebesgue integral
\beao
|P| E[|\rho^l-\ov\rho^l|_T^*] = \sum_{k\in P} E[|\rho^k-\ov\rho^k|_T^*]
\le K \int_0^T  ds \Big\{ \sum_{k \in P}E\Big[\Big| \frac{1}{|C|} \sum_{i\in C} (\rho^i_s - E[\ov\rho_s^C])   -  (\rho_s^k     -\ov\rho_s^k  )\Big|\Big]\Big\}.
\eeao
Now we estimate for $l\in P$ using the triangular inequality}
\beao
E[|\rho^l-\ov\rho^l|_T^*] 
&\le & K \int_0^T  ds \Big\{E\Big[ \Big|\frac{1}{|C|} \sum_{i\in C} (\rho^i_s - E[\ov\rho_s^C])\Big|\Big]   
+ E\Big[\frac1{|P|}\sum_{k\in P} \big|\rho_s^k     -\ov\rho_s^k\big| \Big]\Big\}\\
&=&  K \int_0^T  ds \Big\{E\Big[\Big| \frac{1}{|C|} \sum_{i\in C} (\rho^i_s - E[\ov\rho_s^C])\Big|\Big]   
+ E\big[\big|\rho_s^l  -\ov\rho_s^l \big|\big]\Big\}.
\eeao
Hence, the structure of this inequality is of the form ready to apply Gronwall's Lemma, which yields
\beam\label{boundP}
E[|\rho^l-\ov\rho^l|_T^*] 
&\le & K  \int_0^T  ds \,  E\Big[\Big| \frac{1}{|C|} \sum_{i\in C} (\rho^i_s - E[\ov\rho_s^C])\Big|\Big].
\eeam
Next note that for $i\in C$
\beao
\rho_t^i-\ov \rho_t^i &=& \int_0^t  ds \Big\{ \frac{1-\eps}{|C|-1} \sum_{j\in C\setminus{\{i\}}} \rho^j_s - (1-\eps) \rho_s^i    + (1-\eps) (\ov\rho_s^i- E[\ov\rho_s^C]) \\
&&  + \frac{\eps}{|P|} \sum_{k\in P} \rho^k_s   - \eps\rho_s^i + \eps(\ov\rho_s^i- E[\ov\rho_s^P]) \Big\}.
\eeao
We take all terms under the integral corresponding to the core banks and obtain (we dropped the factor $1-\eps$)
\beao
&& \frac{1}{|C|-1} \sum_{j\in C\setminus{\{i\}}} \rho^j_s - \rho^i_s    + (\ov\rho_s^i- E[\ov\rho_s^C]) \\
&=& \frac{1}{|C|-1} \sum_{j\in C\setminus{\{i\}}} \Big\{
\big(\rho^j_s - \ov\rho_s^j \big) -\big(\rho^i_s- \ov\rho^i_s\big)
 + \big( \ov\rho^j_s - \ E[\ov\rho_s^C]\big)\Big\}.
\eeao
Then we take all terms under the integral corresponding to the periphery banks (dropped the factor $\eps$) and obtain
\beao
&&\frac1{|P|} \sum_{k\in P} \rho^k_s  -\rho_s^i  +  \big( \ov\rho^i_s -  E[\ov\rho_s^P] \big)  \\
&=&  \frac1{|P|} \sum_{k\in P} \left\{
\big(\rho^k_s - \ov\rho_s^k \big) -\big( \rho^i_s-\ov\rho^i_s\big)
  + \big( \ov\rho^k_s- \E[\ov\rho_s^P] \big)
\right\}.
\eeao
Now we estimate for $i\in C$, taking the modulus under the Lebesgue integral and use the triangular inequality
\beao
&& E[|\rho^i-\ov \rho^i |^*_T] \\
&\le & K  \int_0^T  ds 
\left\{\frac{1-\eps}{|C|-1} \sum_{j\in C\setminus{\{i\}}} \Big( E[ |\rho_s^j-\ov \rho_s^j| ]
+ E[|\rho_s^i-\ov \rho_s^i |] \Big) 
+E\left[\Big|\frac{1-\eps}{|C|-1} \sum_{j\in C\setminus{\{i\}}} (\ov\rho_s^j- E[\ov \rho_s^C]) \Big| \right]\right.\\
&& \left. + \frac{ \eps}{|P|} \sum_{k\in P} \Big( E[ |\rho_s^k-\ov \rho_s^k|]
+ E[|\rho_s^i-\ov \rho_s^i |] \Big)+ E\left[\Big|\frac{ \eps}{|P|} \sum_{k\in P}  (\ov\rho_s^k- E[\ov \rho_s^P]) \Big|\right]
\right\}.
\eeao
Summing the previous inequality over all $i\in C$, which are identically distributed, as well as all $\ov\rho_k$ for $k\in P$, 
\beao
\lefteqn{|C| E[|\rho^1-\ov \rho^1 |^*_T]  = \sum_{i\in C} E[|\rho^i-\ov \rho^i |^*_T] }\\
& \le &  K  \int_0^T  ds\left\{ (1-\eps)  \sum_{i\in C}  E[|\rho_s^i-\ov \rho_s^i |] 
+ (1-\eps)  E\left[\Big|\sum_{j\in C\setminus{\{i\}}} \big( \ov\rho_s^j-  E[\ov \rho_s^C]\big) \Big|\right]
 \right.\\
&& \left.+ \eps \sum_{i\in C} \Big( \frac1{|P|}\sum_{k\in P}E[|\rho_s^k-\ov \rho_s^k |]   + E[|\rho_s^i-\ov \rho_s^i |] \Big) + 
\eps E\left[\Big|\frac{|C|}{|P|} \sum_{k\in P}  (\ov\rho_s^k- E[\ov \rho_s^P]) \Big|\right]
\right\}\\
&=&  K  \int_0^T  ds\left\{ (1-\eps)  \sum_{i\in C}  E[|\rho_s^i-\ov \rho_s^i |] 
+ (1-\eps)  E\left[\Big|\sum_{j\in C\setminus{\{i\}}} \big( \ov\rho_s^j-  E[\ov \rho_s^C]\big) \Big|\right]
 \right.\\
&& \left. + \eps \sum_{i\in C} E[|\rho_s^i-\ov \rho_s^i |]  + 
\eps E\left[\Big|\frac{|C|}{|P|} \sum_{k\in P}  (\ov\rho_s^k- E[\ov \rho_s^P]) \Big|\right]
\right\} + K \eps \sum_{i\in C} \int_0^T ds \left\{ \frac1{|P|}\sum_{k\in P}E[|\rho_s^k-\ov \rho_s^k |] \right\} 
\eeao
To estimate the last integral we use the fact that for $k\in P$ all expectations are equal, and take under the integral the supremum over all $s\in [0,T]$. This gives {for arbitrary $l\in P$}
\beao
\int_0^T ds \frac1{|P|}\sum_{k\in P}E[|\rho_s^k-\ov \rho_s^k |] 
& \le & T E[|\rho^l-\ov\rho^l|_T^*] 
\, \le \,  T K \int_0^T ds E \Big[\Big| \frac{1}{|C|} \sum_{i\in C} (\rho^i_s - E[\ov\rho_s^C])\Big|\Big],
\eeao
where the last inequality follows from the bound in \eqref{boundP}.
Now we take this term back under the common integral, recall that all our bounds depend on $T$ and call $TK$ again simply $K$.
Then adding and subtracting $\ov\rho^i_s$ in the above bound, and using the triangular inequality,
\beao
\lefteqn{|C| E[|\rho^1-\ov \rho^1 |^*_T]}\\
& \le &  K \int_0^T  ds \left\{ \sum_{i\in C}   E[|\rho_s^i-\ov \rho_s^i |]  
+  E\left[ \Big|\sum_{i\in C} (\ov\rho_s^i- E[\ov \rho_s^C]) \Big|\right]+  E\left[\Big|\frac{|C|}{|P|} \sum_{k\in P} (\ov\rho_s^k- E[\ov \rho_s^P])\Big|\right]\right\}.
\eeao
This implies
\beam\label{upperbound}
\lefteqn{
 E[|\rho^1-\ov \rho^1 |^*_T]  = \frac1{|C|}\sum_{i\in C} E[ |\rho^i-\ov\rho^i|_T^*]}\\
& \le & K \int_0^T  ds\left\{    E[|\rho_s^1-\ov \rho_s^1 |] 
+ E\left[\Big|\frac{1}{|C|}  \sum_{i\in C}\big(\ov\rho_s^i- E[\ov \rho_s^C]\big)\Big|\right] 
+ E \left[\Big|\frac{1}{|P|} \sum_{k\in P} \big(\ov\rho_s^k- E[\ov \rho_s^P]\big) \Big|\right]
\right\},\nonumber
\eeam
hence, by Gronwall's Lemma, 
\beam\label{eq:Gronwallbound}
\lefteqn{E[|\rho^1-\ov \rho^1 |^*_T] }\\
& \le & K  \int_0^T  \left(
 E\left[\Big|\frac{1}{|C|}  \sum_{i\in C} (\ov\rho_s^i- E[\ov \rho_s^C]) \Big| \right]
+  E\left[\Big|\frac{1}{|P|} \sum_{k\in P} (\ov\rho_s^k- E[\ov \rho_s^P]) \Big|\right]
\right) ds.\nonumber
\eeam
Now we have for $i,j\in V$, since all $\ov\rho^i,\ov\rho^j$ are independent,
\beao
cov(\ov\rho_s^i, \ov\rho_s^j) =  {E[(\ov\rho_s^i-E[\ov\rho_s^{C/P}])(\ov\rho_s^j-E[\ov\rho_s^{C/P}])]} = 0,
\eeao
so that by the Cauchy-Schwarz inequality, for all $s\in [0,T]$,
\beam\label{square}
\Big(E\Big[ \Big | \frac1{|C|} \sum_{i\in C}  \big(\ov\rho_s^i-E[\ov\rho_s^C]\big)
\Big|\Big] \Big)^2
& \le & \frac{1}{|C|^2} E\Big[\Big( \sum_{i\in C} (\ov\rho_s^i-E[\ov\rho_s^C]) \Big)^2\Big]\nonumber \\
 = \,
\frac{1}{|C|^2} \sum_{i\in C} E\Big[ (\ov\rho_s^i-E[\ov\rho_s^C])^2 \Big]
&\le & \frac{1}{|C|} K_C^2,
\eeam
where $K^2_C$ does not depend on $|C|$.
The same argument applies for the sum over $P$, so that we obtain from \eqref{eq:Gronwallbound}
\beam\label{Cbound}
E[|\rho^1-\ov \rho^1 |^*_T] 
 \le   K \left(\frac{1}{\sqrt{|C|} }K_C + \frac{1}{\sqrt{|P|}} K_P\right) <\infty.
\eeam
{For $l\in P$ we go back to \eqref{boundP} and, invoking \eqref{square} {and \eqref{Cbound}}, we find
\beao
E[|\rho^l-\ov\rho^l|_T^*] &\le & K\Big(E[|\rho^1-\ov\rho^1|_T^*] + \int_{0}^{T}ds\Big\{  E\Big[ \Big | \frac1{|C|} \sum_{i\in C}  \big(\ov\rho_s^i-E[\ov\rho_s^C]\big)
\Big|\Big]   \Big\} \Big)\\
& \le & K \left(\frac{2}{\sqrt{|C|} }K_C + \frac{1}{\sqrt{|P|}} K_P\right).
\eeao
 This implies the result.}
\eproof

\brem
(1) \, Note that in the limit system \eqref{limitC} and \eqref{limitP} all processes are independent, so that we have propagation of chaos, meaning that for the system size getting large, all robustness processes become independent.\\
(2) \, From the result \eqref{eq:meanfieldlimit} we see that all banks are mean reverted to a mean process provided that the number of core banks gets large, and the number of core banks and periphery banks satisfy a certain growth condition. In a real market we would think of many more periphery banks than core banks, so that $|C|/|P|\to 0$ as $N\to\infty$ would seem realistic. 
\erem

\section{Risk management in the core-periphery market}\label{s4}

In order to study certain diversification effects in the core-periphery bank model we introduce, similar to \cite{FS,GPY,garnieretal}, friction parameters $\theta_C, \theta_P>0 $ for the core banks and the periphery banks, respectively. 
Hence, the model \eqref{eq:rhocoreequalweights} and \eqref{eq:rhoperipheryequalweights} is extended to
\beam\label{eq:rhocoreequalweights_theta}
d\rho^i_t &=& \theta_{C}\Big(\frac{1-\eps}{|C|-1}\sum_{j\in C\setminus{\{i\}}} \rho^j_t + \frac{\eps}{|P|}\sum_{k\in P} \rho^k_t - \rho_t^i\Big) dt + \si_C dL^i_t,\quad i\in C,\\
d\rho^k_t &=& \theta_P \Big(\frac{1}{|C|}\sum_{i\in C} \rho^i_t  - \rho_t^k\Big) dt + \si_P dL^k_t,\quad k\in P.\label{eq:rhoperipheryequalweights_theta}
\eeam

\bco
The conclusions of Theorem \ref{limitC} hold true  for the extended model \eqref{eq:rhocoreequalweights_theta}  and \eqref{eq:rhoperipheryequalweights_theta} with corresponding limit dynamics
\beam\label{frictionlimitC}
 d\ov\rho^i_t
&=& \theta_C \Big( (1-\eps) E[\ov\rho_t^C] +\eps  E[\ov\rho_t^P] -\ov\rho^i_t\Big) dt + \si_C dL^i_t,\quad i\in C,\\
d\ov\rho^k_t &=& \theta_P \Big( E[\ov\rho_t^C]-\ov\rho^k_t\Big)dt + \si_P dL^k_t,\quad k\in P.\label{frictionlimitP}
\eeam
\eco

We consider $\theta_C$ and $\theta_P$ as parameters emphasizing how strong the corresponding agent is weighting interbank activity in its investment strategy; i.e., a higher value indicates a larger investment into interbank credits. 
This higher value will increase the effect of the mean reversion term in the Ornstein-Uhlenbeck dynamic.

We start the discussion by a simulation study based on  Eq.~\eqref{eq:moup} for the finite network with the specific structure assumed in Theorem~\ref{meanfieldlimit} and the additionally introduced friction parameters; that is
\begin{eqnarray}\label{eq:friction}
d \rho_t = \Theta(A^w - I_{N \times N})\rho_t d t +  \Sigma dL_t,\quad t\ge0,
\end{eqnarray}
where $\Theta$ and $\Sigma$ are diagonal matrices with diagonals $\theta=(\theta_C,\dots,\theta_C,\theta_P,\dots,\theta_P)$ and $\sigma=(\sigma_C,\dots,\sigma_C,\sigma_P,\dots,\sigma_P)$, respectively, for positive constants $\theta_C, \theta_P, \sigma_C$ and $\sigma_P$. 
For our simulation we choose $N = 55$ as network size with $|C| = 5$ and $|P| = 50$. 
For all our simulations the robustness processes { start in 1 and are driven by Brownian motions. }
{Based on data in \cite{craig:2014}} we take $\eps = 0.58$.

\subsection{Hedging changes in the market volatility}\label{sec:sim}

We examine the consequences of changes in the market volatilities to either core or periphery banks, based on the paths of the agents' robustnesses.
We consider different scenarios.

\begin{figure}[t]
\begin{center}
\includegraphics[width=1\textwidth, height=0.35\textheight]{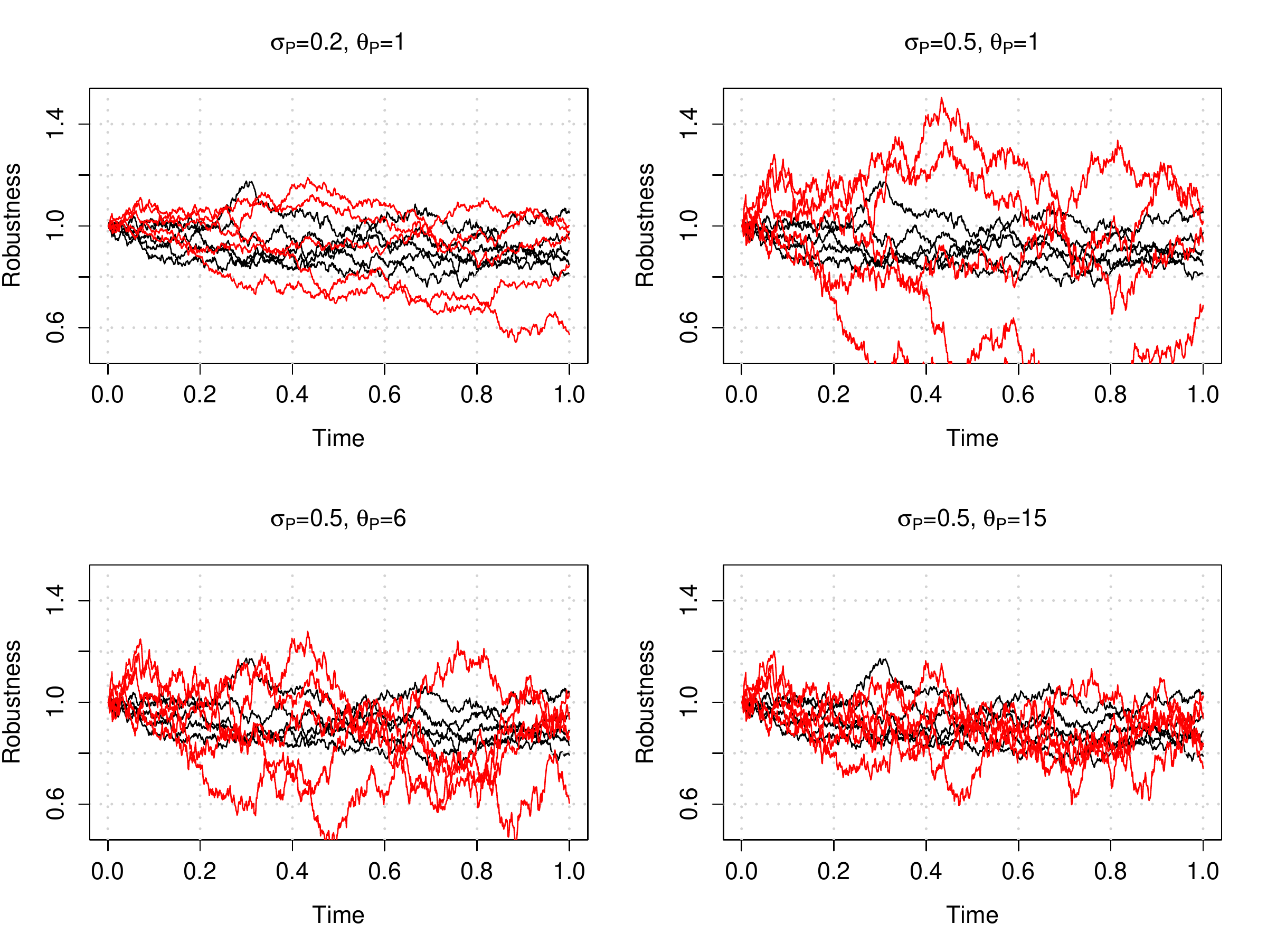}\\[-2cm]
\end{center}
\vspace*{-1cm}
\caption{\label{plot1}  Realized robustness processes of a simulation on the core-periphery banking network as described in the text. 
The robustness of the five core banks are depicted in black and five (out of 50) periphery banks in red.
All four plots are {simulated with the same random seed}, but differ due to varying parameters $\sigma_P$ and $\theta_P$. 
In all four plots $\sigma_C=0.2$ and $\theta_C=1$ remain constant. }
\end{figure}

Initially core and periphery will face the same economic environment and we choose $\sigma_C = \sigma_P = 0.2$ and $\theta_C = \theta_P = 1$, respectively. 
For this choice of parameters Figure~\ref{plot1} shows in the upper left plot sample paths of the robustness for all five core banks and five (out of 50) periphery banks.
In a next step we suppose higher volatility in the market. 
Whereas core banks can keep the volatility of their non-interbank assets constant due to sophisticated hedging strategies; i.e., the same value $\sigma_C = 0.2$ holds,  periphery banks do not have the resources and expertise for such methods. 
Hence, the standard deviation of their non-interbank assets increases to $\sigma_P = 0.5$. 
If periphery banks do not undertake a shift of their assets but keep their investment strategy unchanged, their robustness will show a higher variation, which is confirmed by the upper right plot  in Figure~\ref{plot1}. 
The two lower plots in the same figure highlight that an increase of $\theta_P$ (by increased investment into interbank credits) can reduce variation.  

A further analysis of the consequences of such a change in the market volatility is summarized in Table \ref{tab1}, where $\hat \rho_1^P$ is the estimated robustness of one periphery bank at time $t=1$ based on 100 simulation runs for varying parameters of $\theta_P$. 
The standard errors in brackets indicate that an increase of $\theta_P$ can indeed reduce the variation in the robustness of a periphery bank.

So far, our model suggests that periphery banks can reduce uncertainty in their robustness, resulting from higher volatility in non-interbank assets, by higher investment into interbank assets. 
This strategy has certain drawbacks, when a shock hits all core banks' robustness at the same instant of time. 
For pointing out the resulting effect we redo the simulations and assume a reduction of all core banks' robustness by 0.3 at $t = 0.9$ and investigate the market at $t=1$ (immediately after the shock) and at $t=2$.
Such a shock, being restricted to the core, means that for a short term the robustness of core banks and periphery banks will diverge. 
However, due to the mean reversion in Eq.~\eqref{eq:friction} the mean of core and periphery banks' robustnesses will again revert to a common value in the long run (cf. Corollary~\ref{OUstationary} below). 
Table~\ref{tab1} illustrates that in a shock scenario an increased $\theta_P$ still reduces variation, but a shocked core will affect the periphery more intensive for higher values of $\theta_P$. 
For smaller values of $\theta_P$ the robustness of the periphery banks exhibits a lower sensitivity with respect to the shock on the core. In this case core banks can in turn benefit from their interbank activity with more robust periphery banks. 
This becomes apparent in the estimates of the robustness at $t = 2$, which  show approximately the new common robustness of core and periphery in the post-shock regime. 
Apparently, for $\theta_P = 1$ the core banks can at least recover partially from the shock, which is, however, not the case any more, if $\theta_P$ becomes too large. 
 
Overall, we conclude that periphery banks can have an incentive to invest more into the interbank market in order to hedge their volatility, however, the increase of interbank investment makes them more vulnerable for contagion resulting from a core-wide shock. 
The whole network will also suffer, if the periphery invests too much into the core as the increased sensitivity of the periphery with respect to the core's constitution will have negative feedback effects on the core itself and its ability to recover from past shock events. 

\begin{table}[t]
\begin{small}
\begin{center}
\begin{tabular}{ c||c|c||c|c|c|c|| }
  
  \multicolumn{1}{c||}{}&\multicolumn{2}{|c||}{without shock {($t = 1$)} } &\multicolumn{2}{|c|}{with shock ($t = 1$)}&\multicolumn{2}{|c||}{with shock ($t = 2$)} \\
  \hline\hline
  $\theta_P$ & $\hat \rho_1^P$ & $\hat \rho_1^C$& $\hat \rho_1^P$ & $\hat \rho_1^C$ &$\hat \rho_2^P$ & $\hat \rho_2^C$ \\
  \hline\hline
  $1$ &0.99 (0.34) &0.98	(0.14)&0.96 (0.34) &0.70 (0.14)	&0.77 (0.33)&0.75 (0.15) \\
\hline
  $3$ & 1.00 (0.22) & 0.98 (0.14)&0.92 (0.22)&0.69 (0.14)& 0.71 (0.24)&0.71 (0.15) \\
\hline
  $6$ & 1.00 (0.16) & 0.98 (0.14)& 0.86 (0.16)&0.69 (0.14)& 0.70 (0.19)&0.70 (0.15)\\
\hline
  $10$ &1.00 (0.14) &0.98 (0.14)&0.81 (0.14)&0.69 (0.14)&0.69 (0.16)&0.69 (0.16)\\
\hline
  $15$ & 1.00 (0.13) &0.98 (0.14)&0.77 (0.13)&0.69 (0.14)&0.69 (0.15)&0.69 (0.16)\\
\hline
  $20$ & 1.00 (0.12)&0.98 (0.14)&0.74 (0.12) &0.69 (0.14) &0.69 (0.14)&0.68 (0.16)\\
\hline
  $25$ & 1.00 (0.11)& 0.98 (0.14)&0.72 (0.11)&0.69 (0.14)&0.69 (0.14)&0.68 (0.16) \\
\end{tabular}
\end{center}
\end{small}
\caption{\label{tab1} Estimated robustness of one core and one periphery bank, respectively.
Presented are the empirical means based on 100 simulation runs with standard errors in brackets. 
The figures are based on a scenario of increased volatility; i.e., an increase from $\sigma_P = 0.2$ to $\sigma_P = 0.5$. 
The simulation for the original volatility and mean reversion, $\sigma_P = 0.2$ and $\theta_P = 1$, results in the mean estimate {$\hat\rho_1^P=0.99$} with standard error  0.14. 
Thus, in the regime of higher market volatility a periphery bank can approximate the magnitude of variation from the previous regime of lower volatility by setting $\theta_P = 10$. }
\end{table}

\subsection{Risk management of structural breaks in the market}\label{sec:fpt}

In this section we want to shed light on the outcome of the previous simulations in a more concrete way by relying on first passage times. 

The following corollary presents the first and second moments of the limit processes. 

\bco\label{OU-meanmodel}
The SDEs \eqref{frictionlimitC} and \eqref{frictionlimitP} of the limit model have independent Ornstein-Uhlenbeck dynamics with solutions
\beao
\ov\rho^i_t &=&  e^{-\theta_C t} \; \ov\rho^i_0 + \theta_C
\int_0^t e^{-\theta_C(t-u)} ((1-\eps) E[\ov\rho^C_u]+ \eps  E[\ov\rho^P_u]) du + \si_C \int_0^t e^{- \theta_C(t-u)} dL^i_u,\quad i\in C,\\
\ov\rho^k_t &=&  e^{-\theta_P t}  \; \ov\rho^k_0 + \theta_P
\int_0^t e^{-\theta_P (t-u)} E[\ov\rho^C_u] du + \si_P \int_0^t e^{-\theta_P (t-u)} dL^k_u,\quad k\in P,
\eeao
Moreover, $\ov\rho^i_t$  for $ i=1,\ldots, N$ have second order moment structure 
\beao
E[\ov\rho^i_t\mid \ov\rho^i_0=a_i] &=& e^{-\theta_C t}a_i+\theta_C \int_0^t e^{-\theta_C (t-u)}  \left((1-\eps) E[\ov\rho^C_u] + \eps  E[\ov\rho^P_u]\right) du,\quad i\in C, \\
E[\ov\rho^k_t\mid \ov\rho^k_0=a_k] &=&   e^{-\theta_P t}a_k+\theta_P \int_0^t e^{-\theta_P (t-u)}  E[\ov\rho^C_u] du,\quad  k\in P,\\
 \var[\ov\rho^i_t\mid \ov\rho^i_0=a_i] &=& \frac{\si_C^2}{2\theta_C}(1-e^{-2 \theta_C t}),\quad i \in C\\
 \var[\ov\rho^k_t\mid \ov\rho^k_0=a_k] &=& \frac{\si_P^2}{2\theta_P}(1-e^{-2 \theta_P t}),\quad k \in P\\
Cov[\ov\rho^i_s,\ov\rho^i_t\mid \ov\rho^i_0=a_i] &=&\frac{\si_C^2}{2\theta_C }(e^{-\theta_C |s-t|}-e^{-\theta_C (s+t)}),\quad i \in C \\
Cov[\ov\rho^k_s,\ov\rho^k_t\mid \ov\rho^k_0=a_k] &=&\frac{\si_P^2}{2\theta_P }(e^{-\theta_P |s-t|}-e^{-\theta_P (s+t)}),\quad k \in P.
\eeao
\eco

Provided there exists a stationary version of the system of Corollary~\ref{OU-meanmodel}, 
this stationary model has constant means and variances of the core and periphery banks, respectively.
They are obtained from the above moments for $\tto$, which yields $E[\ov\rho^C]=E[\ov\rho^P]=:\mu$ and 
$\var[\ov\rho^i]=\si_C^2/(2\theta_C)$ for $i\in C$ and $\var[\ov\rho^k]=\si_P^2/(2\theta_P)$ for $k\in P$.
The resulting stationary dynamics lead to a further simplification of the original model.

\bco\label{OUstationary}
Stationary versions of the SDEs in Corollary~\ref{OU-meanmodel} are given by
\beao\label{statOU}
d\ov\rho^i_t &=& \theta_C \big( \mu -\ov\rho^i_t\big)dt + \si_C dL^i_t,\quad i\in C,\\
d\ov\rho^k_t &=& \theta_P \big( \mu -\ov\rho^k_t\big)dt + \si_P dL^k_t,\quad k\in P,
\eeao
where $\mu$ is the a.s. limit of the mean robustness of the core banks $\frac1{|C|}\sum_{i\in C} \rho^i_t$ as $|C|\to\infty$ for all $t\ge 0$. 
\eco

Consequently, for a large number of core and periphery banks, we can discuss various risk measures by relying on the simple Ornstein-Uhlenbeck dynamic of Corollary~\ref{OUstationary}.
As a first risk measure we consider the standard deviation.

\bde
For each bank we define the {\em standard deviation risk}
$$S_i={\sqrt{\si_C^2/(2\theta_C)}}\ \text{for}\ i\in C\quad\text{and} \quad  S_k={\sqrt{\si_P^2/(2\theta_P)}}\ \text{for}\ k\in P. $$
\ede

It is certainly one goal of every bank $i\in V$ to keep $S_i$ within certain bounds and at best constant (cf. \cite{garnieretal}). 
As a second risk measure we define the inverse of the mean first passage time of the robustness of an agent.  

\bde\label{def:ifpt}
For every bank $i\in V$ denote the {\em inverse first passage time risk}  (IFPT risk) by
$$\tau_i = 1/E[T_i(0)]$$ 
where $T_i(0):=\inf\{t\ge 0 : \ov\rho^i_t=0\}$ denotes the first passage time of $\ov\rho^i$ to 0.
\ede

First passage events have also served as triggering events to start a cascading mechanism in the market (e.g. Battiston et al.~\cite{battiston:2012}).
For a mathematical analysis \cite{battiston:2012} approximates the first passage time of a mean reverting OU process simply by that of Brownian motion. We prefer to work with the following precise formula.

\ble
Assume the core-periphery bank system driven by independent Brownian motions and assume that all robustness processes are solutions to the SDEs in Corollary~\ref{OUstationary} with starting values $\ov \rho^i_0=1$ for all $i\in V$.
Define $\al:=\si/\sqrt{2\theta}$ and let $\ov\Phi=1-\Phi$ denote the tail and $\vp$ the density of the standard normal distribution.
Let $T(0)$ denote the first passage time of a generic bank $\ov\rho$ to 0.
Then the following hold:
\beam\label{fht}
E[T(0)] &=& 2\, \frac{\al}{\si^2} \int_0^1  \frac{\ov \Phi(  \frac{v-\mu}{\al} ) }{\vp( \frac{v-\mu}{\al}) }dv
 = \frac1{\theta} \int_{-\mu/\al}^{(1-\mu)/\al}  \frac{\ov\Phi(y)}{\vp(y)} dy.
\eeam
\ele

\bproof
According to Prop.~4 of \cite{WardGlynn}  the first passage time of an Ornstein-Uhlenbeck process 
$$dX_t = \theta(\mu-X_t) dt + \si dW_t$$
which starts in 1 to hit 0 for the first time has expectation
\beam\label{expfpt}
E[T(0)] = \sqrt{\frac{4\pi }{\theta\si^2}} \int_0^1 \exp\left\{\Big( \frac{v-\mu}{\si}\Big)^2\theta\right\}
P\Big(N\big(\mu,\frac{\si^2}{2\theta }\big)>v\Big) dv,
\eeam
where $N(\cdot,\cdot)$ is a normal random variable with mean and variance in the first and second component. 
Then, setting $\al:=\si/\sqrt{2\theta}$, we get
\beam
E[T(0)] &=& \sqrt{\frac{4\pi }{\theta\si^2}} \int_0^1 \exp\left\{\Big( \frac{v-\mu}{\si}\Big)^2\theta\right\}
P\Big(\mu+ \sqrt{\frac{\si^2}{2\theta }} N(0,1)>v\Big) dv\nonumber\\
&=& \sqrt{\frac{4\pi }{\si^2}} \int_0^1 \frac1{\sqrt{\theta}}\exp\left\{\Big( \frac{v-\mu}{\si}\Big)^2\theta\right\}
\ov \Phi\Big(\sqrt{2\theta}\frac{v-\mu}{\si} \Big) dv\nonumber\\
&=& \frac{2\al }{\si^2} \int_0^1 \sqrt{2\pi} \exp\left\{\frac{1}{2}\Big( \frac{v-\mu}{\al}\Big)^2\right\}
\ov \Phi\Big(\frac{v-\mu}{\al} \Big) dv.\label{fptalpha}
\eeam
so that, since $\exp\{- y^2/2\}=\sqrt{2\pi}\vp(\sqrt{2\theta}y)$, the integrand can be rewritten as
\beao
f\Big(\frac{v-\mu}{\al}\Big) := 
 {\ov  \Phi \Big(\frac{v-\mu}{\alpha} \Big) }\Big/{\vp \Big(\frac{v-\mu}{\alpha} \Big)},
\eeao
where a substitution of variables yields the final result.
\eproof

The first passage time $T_i(0)$ of the robustness process of bank $i$ can be interpreted as default of a bank. 
Hence, any bank will surely aim to keep $\tau_i$ low. 

We will asses the risk management of a periphery bank in a more quantitative manner, if structural breaks on the core occur. 
The risk measures of interest are the standard deviation risk $S_P= \sqrt{\sigma_P^2/(2\theta_P)}$ and the IFPT $\tau_P$ for any periphery bank. 

We first come back to the scenario in Section~\ref{sec:sim}, where we have found out that for a bank with fixed target value for $S_P$, when the volatility $\sigma_P$ is varying, the periphery bank can hedge this by choosing an appropriate $\theta_P$.
Now we can quantify this value, namely,
\begin{eqnarray}\label{eq:theta}
\theta_P = \sigma_P^2 / (2S_P^2)
\end{eqnarray}
for any value of $\sigma_P$. 
Particularly, an increase of $\sigma_P$ requires an increase in $\theta_P$, which is in line with the simulations illustrated in Figure~\ref{plot1}. {For $\sigma_P$ changing from 0.2 to 0.5 (cf. Section~\ref{sec:sim}), for instance, Eq.~\eqref{eq:theta} suggests a required increase of $\theta_P$ from 1 to 6.25 in order to keep the standard deviation risk $S_P$ constant at its initial value 0.1414. Without changing $\theta_P$ the periphery bank would have to accept the standard deviation risk $S_P$ rising from 0.1414 to 0.3535 implying a higher uncertainty for the future robustness.}

\begin{figure}[tb]
\begin{center}
\includegraphics[width=0.6\textwidth]{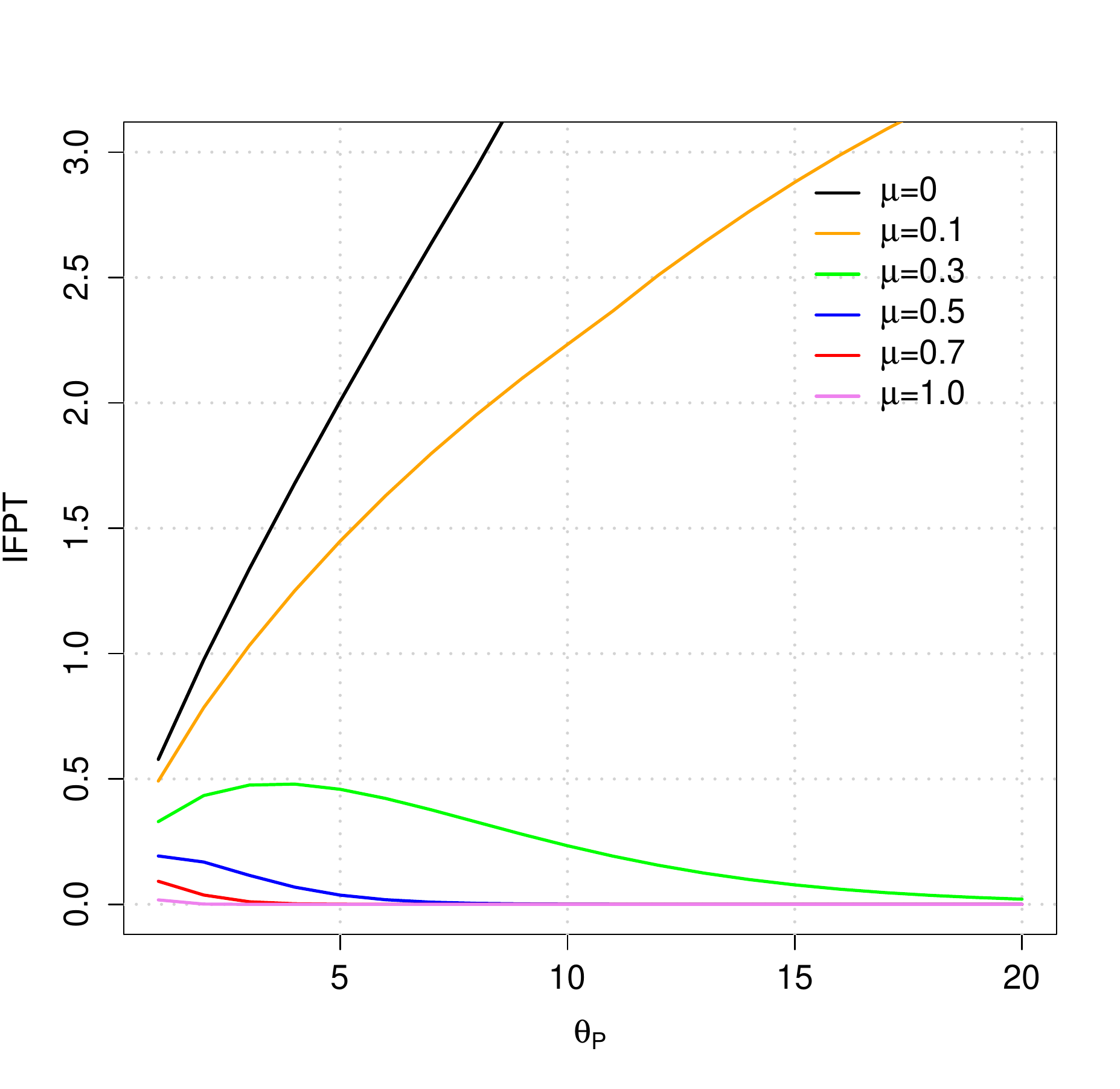}
\end{center}
\vspace*{-1cm}
\caption{\label{plot7} 
IFPT risk  for different values of $\mu$ as a function of $\theta_P$.
}
\end{figure}

As noticed from the values in Table~\ref{tab1} larger values of $\theta_P$, however, may reinforce the decrease of a periphery bank's robustness, if the robustness of core banks is relatively low. {This drawback of increasing $\theta_P$ becomes apparent, if one takes the IFPT risk $\tau_P$ into account. Figure~\ref{plot7} shows how the IFPT risk depends on $\theta_P$ for $\sigma_P = 0.5$ and different values of the mean robustness $\mu$ in the core. A reduction in $S_P$ by enlarging $\theta_P$ has beneficial effects on the IFPT risk only as long as $\mu$ stays sufficiently large. However, for values of $\mu$ near zero an increase of $\theta_P$ results in a massive increase of the IFPT risk.}

\begin{figure}[t]
\begin{center}
\includegraphics[width=0.6\textwidth]{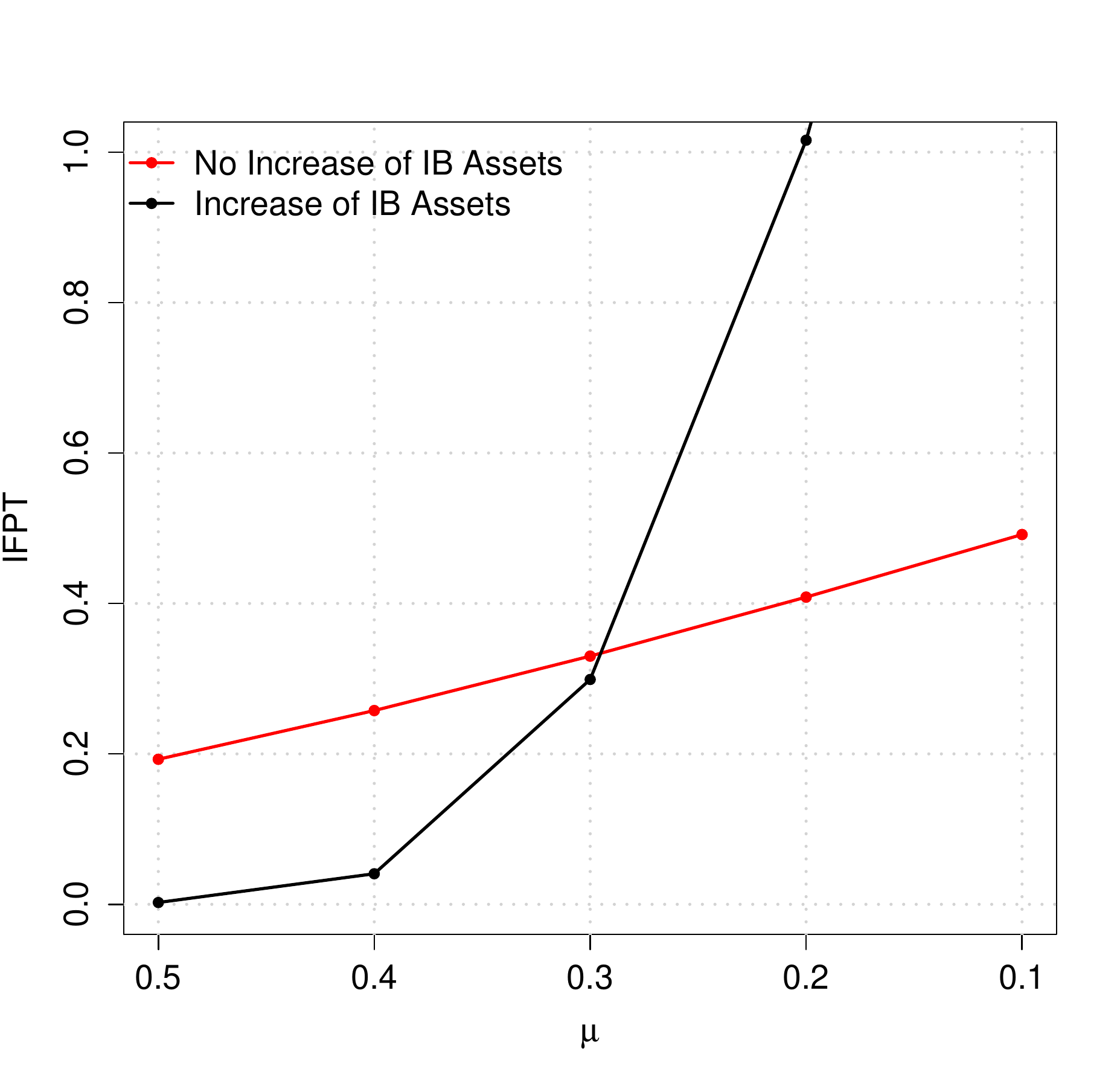}
\end{center}
\vspace*{-1cm}
\caption{\label{plot6} 
IFPT risk for different values of the mean robustness $\mu$ in the core of the interbank market. The figure compares two investment strategies: Increase of interbank assets vs. no increase of interbank assets { in a regime of $\sigma_{C}=0.2$ and  $\sigma_P = 0.5$.} 
}
\end{figure}

We like to substantiate this by a final example which adopts the setting of Section~\ref{sec:sim}; i.e., we assume again an increase of volatility  from $\si_P=0.2$ to $\si_P=0.5$ hitting the periphery. 
Instead of hedging $S_P$ we now aim at keeping the IFPT risk $\tau_P$ constant by adjusting $\theta_P$ correspondingly. 
For $\mu = 0.5$ and the assumed volatility scenarios, we have computed -- by applying a numerical rootfinder to Eq.~\eqref{fht} --  that a periphery bank must increase $\theta_P$ from 1 to 8.6 in order to keep $\tau_P$ constant at the low level of $\tau_P = 0.002$. 
Otherwise, without an adjustment of the interbank investment volume, the IFPT $\tau_P$ would jump to 0.192. 
For an illustration, see Figure \ref{plot6}, which compares $\tau_P$ for both alternatives, increase of $\theta_P$ and no increase of $\theta_P$. 

Similar to the case of hedging $S_P$ previously, the drawback of keeping $\tau_P$ constant arises, if  $\mu$ changes. 
Assume that some structural break within the core occurs, resulting from an external shock, which cuts now the mean robustness of the core banks. 
Figure~\ref{plot6} suggests that the remedy of an increased investment into interbank asset would turn out to be the worse alternative as soon as the core robustness experienced a reduction in the mean robustness of slightly more than 0.2. 
When such an event  happens, the periphery bank's IFPT risk would have been smaller without increasing the interbank investment for hedging the IFPT risk $\tau_P$ under increased volatility.

\section{Conclusion}\label{s5}

Based on empirical evidence we employed a hierarchical block model for modelling the interbank market as a network, in which a small number of highly connected large banks (the core) play the role of financial intermediaries for a large number of smaller banks (the periphery). 
We introduced the financial robustness of the agents as continuous-time stochastic processes explicitly incorporating the market structure.  Further, we proved a LLN for this coupled multivariate system as the network size grows.
 We proved that in the limit system all processes are independent, hence, the system decouples more and more as the network enlarges. 
 This behaviour is called \textit{propagation of chaos} in the physics community.
 
In a first simulation approach on the core-periphery network we have pointed out that risk management decisions, although being meaningful from the perspective of a single agent may accelerate the negative effects of a system-wide distortion. 
Our application is based on the assumption that core banks have more expertise and resources available for performing sophisticated risk management measures in order to hedge volatility successfully on the non-interbank market.
In contrast, the only possibility of periphery banks to hedge increasing volatility occurring in their non-interbank assets portfolio is provided by expanding the investment into interbank assets, that is for instance an increase of deposits at core banks. 
Our model and the chosen risk measures in Section~\ref{sec:fpt} disclose some analytical tools for evaluating the risk management decisions of a periphery bank under such conditions. 
As in previous simulation approaches  we have observed that periphery banks can hedge volatility via increasing interbank investments, however, this at first effective hedging activity may become a drawback in the case of external shocks hitting parts of the network.

In our paper we have established a framework, which gives a basis  for further examination of the interbank market - particularly under the viewpoint of interaction between periphery and core banks. 
Further research will relax the still restrictive core-periphery model from Section~\ref{s3} and  prove an analogue of Theorem~\ref{meanfieldlimit} with off-diagonal blocks in the core-periphery adjacency matrix allowing for a higher degree of heterogeneity by not assuming full lending relationships. 
Also central limit theorems, Poisson limit results and large deviation results will provide further interpretations for systemic risk.

\subsubsection*{Acknowledgement}
We take pleasure in thanking {Carsten Chong,} Ben Craig, Jean-Dominique Deuschel, Nina Gantert, and Daniel Matthes for interesting discussions.
{OK and LR also appreciated discussions with the participants of the Research Seminar of the Deutsche Bundesbank, where LR presented our work.}

\bibliographystyle{plain}
\bibliography{Bib}

\end{document}